\documentclass[conference]{IEEEtran}
\IEEEoverridecommandlockouts
\usepackage{cite}
\usepackage{amsmath,amssymb,amsfonts}
\usepackage{xcolor}
\usepackage{graphicx}
\usepackage{capt-of}
\usepackage{url}

\definecolor{dark-red}{rgb}{0.4,0.15,0.15}
\definecolor{dark-blue}{rgb}{0.15,0.15,0.6}
\definecolor{medium-blue}{rgb}{0,0,0.5}

\usepackage{xcolor}

\long\def\invis#1{}

\newcommand\gderror[1]{
   \typeout{--------------------------------------------------------------------}
   \typeout{------- #1 ---------}
   \typeout{--------------------------------------------------------------------}
   {\bf #1}
}
\newcounter{gdTmp} 
\newcounter{gdLastCount}
\newcommand\maxpage[2][Error]{  
\ifnum\value{page}>#2
    \gderror{On page {\thepage} we are past page #2 (too long).   #1 }
\else\fi
\setcounter{gdLastCount}{\value{page}} 
}
\newcommand\maxpageSinceLast[2][Error]{  
\ifnum \numexpr \value{page} - \value{gdLastCount}\relax>#2
    \gderror{Exceeds max length #2 pages. Page \thepage: #1}
\thepage\else\fi
\setcounter{gdLastCount}{\value{page}} 
}

\def\BibTeX{{\rm B\kern-.05em{\sc i\kern-.025em b}\kern-.08em
    T\kern-.1667em\lower.7ex\hbox{E}\kern-.125emX}}

\makeatletter
\newcommand{\newlineauthors}{%
  \end{@IEEEauthorhalign}\hfill\mbox{}\par
  \mbox{}\hfill\begin{@IEEEauthorhalign}
}
\makeatother

\begin{document}

\title{Do LLMs Beat Nash? Testing Decentralized Coordination in Self-Play Multi-Agent Games}
\author{
\IEEEauthorblockN{Deborah Sinishaw, Qile Zhu, Edwin Meriaux, and Gregory Dudek}
\IEEEauthorblockA{Centre for Intelligent Machines, School of Computer Science, McGill University}
\thanks{D. Sinishaw is with the Department of Electrical and Computer Engineering, McGill University (e-mail: deborah.sinishaw@mail.mcgill.ca). Q. Zhu and E. Meriaux are with the School of Computer Science, McGill University (e-mail: qile.zhu@mail.mcgill.ca; edwin.meriaux@mail.mcgill.ca). All authors are affiliated with the Centre for Intelligent Machines (CIM), McGill University, and are supervised by G. Dudek (e-mail: gregory.dudek@mcgill.ca).}
}
\maketitle

\begin{abstract}
Large language model agents deployed without a central controller are often assumed to require communication to coordinate their actions. We ask what remains possible without it: when independent instances of the same model cannot communicate, can they still reason about their counterparts well enough to exceed the standard game-theoretic baseline for uncoordinated play? We introduce a benchmark of one-shot, no-communication games in which each of thirteen language models is told only that its counterparts are running the same model and is evaluated against the Nash equilibrium of the underlying game. In two-player matrix games spanning seven archetypes and two to ten actions per player, two frontier-hosted models consistently exceed their Nash benchmark, approaching the optimal joint outcome in several archetypes, while most open-weight models achieve only partial gains that vary sharply by game structure. Performance degrades substantially in team-based games with four or more interchangeable agents, particularly as the action space grows, suggesting that whatever capability drives self-play gains in dyadic games does not transfer to larger multi-agent teams.
\end{abstract}

\begin{IEEEkeywords}
Decentralized Coordination, Game Theory, Large Language Models, Multi-agent Systems, Nash equilibrium, Robotics
\end{IEEEkeywords}

\section{Introduction}
\label{sec:intro}         
Coordinating multiple agents without a central controller is a long-standing problem in multi-agent robotics~\cite{cao1997cooperative,gerkey2004taskallocation,ismayilov2025decentralized}, where each agent must choose its actions while accounting for decisions it cannot observe or dictate. Depending on the game, agents may compete or cooperate, but in both cases their payoffs depend on the joint actions of all participants.

LLMs are increasingly deployed as agents~\cite{akata2025repeated,agashe2025coordination}. In practice several agents in a system often run the same underlying model while holding separate contexts and no shared memory. This is a problem for applications such as multi-robot exploration where a communication channel may be unreliable, costly, or simply unavailable at the moment a decision must be made. Prior work evaluates LLM agents in repeated play or with a communication channel, where agents can converge over time or negotiate explicitly.

This leads to our central research question: \emph{which coordination problems can independent LLM agents solve without communication?} Success may depend not only on model capability but also on the strategic structure of the task. Some games require agents to converge on the same action, whereas others require them to break symmetry and assume distinct roles. Success at the first does not necessarily imply success at the second. 

\begin{center}
\begin{minipage}{\columnwidth}
    \centering
    \includegraphics[width=0.85\linewidth]
        {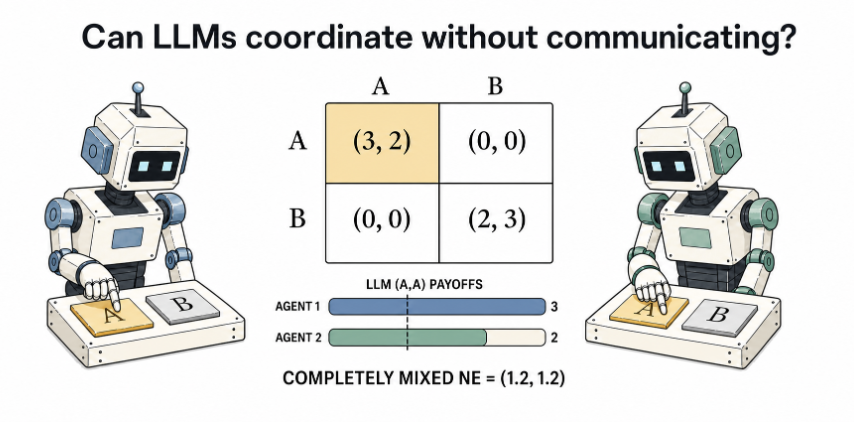}
    \captionof{figure}{Without communication, two LLM agents independently
    select $A$, obtaining $(3,2)$ rather than the mixed-Nash payoff
    $(1.2,1.2)$.}
    \label{fig:bos_example}
\end{minipage}
\end{center}

This paper makes two contributions. First, we provide a parameterized game suite and generator with verified Nash benchmarks across multiple game archetypes. Second, an evaluation of independent copies of the same model across this suite in one-shot play without communication, with
action labels rotated to control for position bias.

\section{Background and Related Work}
\label{sec:background}

\subsection{Game-Theoretic Preliminaries}
In game theoretic settings, a game defines players, actions, and the payoff for each joint action. A Nash equilibrium is a game profile in which no player gains by deviating alone; every finite game has at least one in mixed strategies~\cite{nash1950equilibrium}. Because mixed Nash equilibrium assumes independent randomization, it is the reference point for agents without a shared signal. Correlated equilibrium relaxes this and can yield higher joint
payoffs~\cite{aumann1974correlated}, so scoring above Nash without a channel indicates coordination from something other than an explicit correlating
device. We use Nash benchmarks for both the two-player and team games; scoring is defined in Section~\ref{sec:protocol}.

\subsection{Related Work}
Classical theories explain how agents can coordinate without communicating. Focal points and conventions hold that some option is simply more salient than the alternatives, so independent agents converge on it without needing to interact~\cite{schelling1960strategy,lewis1969convention}. Another work explains that superrationality and program equilibrium instead derive coordination from each agent knowing that the other reasons in exactly the same way. This allows it to predict the other's move by reasoning about itself~\cite{hofstadter1983superrationality,tennenholtz2004program}. Lastly, other-play asks what coordination remains possible when this shared-reasoning assumption cannot be relied on, for instance because conventions or labels differ across agents~\cite{hu2020otherplay}.

Empirical work on LLM agents has focused less on testing this no-communication and no-history setting. Studies of repeated play let agents adapt using the observed outcomes of earlier rounds~\cite{akata2025repeated}, coordination benchmarks typically provide an explicit communication channel~\cite{agashe2025coordination}, and game-theoretic evaluations often combine both~\cite{duan2024gtbench}. In each case, agents have some means of coordinating beyond reasoning alone.

We remove both mechanisms: agents play a single round with no history to learn from and no channel to negotiate through, using identical models on both sides. We then vary game structure, action count, and team size to see what coordination, if any, survives under these conditions.

\section{Methodology}
\label{sec:methodology}
\subsection{Game Suite Design}
Our benchmark combines two game families. Tier~1 uses seven standard two-player matrix games plus three controls (as seen in Table~\ref{tab_1}). Tiers~2 and~3 use six generated team-game archetypes, each a distinct strategic structure, summarized in Table~\ref{tab:archetypes}.\footnote{The game suite and evaluation harness will be released publicly upon acceptance at \url{https://github.com/Dxborah/llm-nash-coordination}.}

\begin{table}[t]
\caption{Two-player game archetypes (Tier 1).}
\label{tab_1}
\centering
\footnotesize
\begin{tabular}{p{0.19\columnwidth}p{0.52\columnwidth}p{0.19\columnwidth}}
\hline
\textbf{Game} & \textbf{Mechanism} & \textbf{Status} \\
\hline
Coordination & Both players gain by choosing the same action. & Scored \\
Stag hunt & Mutual cooperation pays most but is risky; the safe action guarantees less. & Scored \\
Chicken & Each prefers to hold firm, but mutual aggression is the worst outcome. & Scored \\
Prisoner's dilemma & Defection dominates, yet mutual cooperation beats mutual defection. & Scored \\
Public goods & Both share the benefit of contribution, but contributing is privately costly. & Scored \\
Anti-coordination & Both players gain by choosing different actions. & Scored \\
Battle of the sexes & Both want to match, but prefer different equilibria. & Scored \\
Matching pennies & One player wins when the actions match, the other when they differ. & Control \\
Cyclic zero-sum & Rock-paper-scissors: each action beats some and loses to others. & Control \\
Dominance-control & One action strictly dominates all others. & Control \\
\hline
\end{tabular}
\end{table}

\begin{table}[t]
\caption{Generated team-game archetypes (Tiers 2--3).}
\label{tab:archetypes}
\centering
\footnotesize
\begin{tabular}{p{0.19\columnwidth}p{0.52\columnwidth}p{0.19\columnwidth}}
\hline
\textbf{Archetype} & \textbf{Mechanism} & \textbf{Status} \\
\hline
Zero-sum & One team gains when the other team loses. & Zero-sum \\
Coordination & Both teams get more when their action shares are close. & General-sum \\
Threshold & Both teams get a reward only if each reaches its own target count. & General-sum \\
Public goods & All agents share the benefit of effort, but own effort has a cost. & General-sum \\
Best shot & The highest effort can help both teams, but own effort has a cost. & General-sum \\
Congestion & Both teams get less when the combined load on a resource grows. & General-sum \\
\hline
\end{tabular}
\end{table}

\subsection{Composition-Based Payoff Representation}
For tiers 2--3, teams contain two or three agents. Rather than the individual action profile of each player, we track only each team's \emph{composition}: the count of agents choosing each action. This is valid because team members are interchangeable in the payoff rule, and it sharply reduces payoff-matrix size. For a team of $n$ agents with $k$ actions, the number of distinct compositions is $\binom{n+k-1}{k-1}$ rather than $k^n$ profiles: with two actions, $n+1$ counts instead of $2^n$; with three, $\binom{n+2}{2}$ (6 for $n=2$, 10 for $n=3$). Two three-agent teams over three actions thus face a $10\times10$
payoff matrix instead of $27\times27$, keeping games tractable while preserving strategic structure.

\subsection{Game Generation and Nash Computation}
We use non-round parameter values throughout to reduce accidental payoff ties. Pure Nash equilibria are found by exhaustive search over payoff-matrix cells and mixed Nash equilibria are computed with the Lemke--Howson algorithm~\cite{lemkehowson1964}. Exact payoff ties can make a game degenerate and stall the solver, so for the three-action suite we add a small random perturbation to payoffs only during solving, then discard it and report the mixed solution against the original payoffs. Both pure and mixed equilibria are stored as benchmark values for the LLM evaluation.

\section{Experimental Protocol}
\label{sec:protocol}
The benchmark has three tiers. They are all one-shot self-play: independent copies of a model that act simultaneously with independent seeds, no communication, and no shared memory. Tier~1 is a two-player matrix suite ($k=2$--$10$ actions). Tiers~2 and~3 regroup play into the six generated team archetypes of Table~\ref{tab:archetypes}, with interchangeable agents forming two teams and two (Tier~2) or three (Tier~3) actions each. Every tier cyclically rotates the action labels and rotates responses back before scoring. This prevents a model favoring the first-listed option from appearing to coordinate; the spread of a game's score across rotations is its \emph{label\_bias}. Given this setup roughly $53,000$ tests are run for the sake of this paper. Sampling temperature is $0.7$ for the two-player suite (GPT-5.6 uses its fixed default); the team tiers use the backend default.

\textbf{Tier 1.}\label{sec:2p-protocol} Each trial queries two model instances in
parallel as Players~1 and~2. Each is shown the payoff table and told to maximize its
own reward against the other player, and returns a schema-enforced JSON distribution
over its $k$ actions plus an intended action; the two strategies give the expected
joint payoff. The suite defines seven scored games and three controls
(Table~\ref{tab_1}). Each payoff is normalized to a two-sided score,
\begin{equation}
s =
\begin{cases}
\dfrac{a - N}{\,O - N\,}, & a \ge N,\\[8pt]
\dfrac{a - N}{\,N - W\,}, & a < N,
\end{cases}
\label{eq:efficiency}
\end{equation}
where $a$ is the achieved joint payoff, $N$ the worst (lowest-welfare) symmetric
equilibrium, $O$ the optimal (best joint) cell, and $W$ the worst cell. Upside and
downside are normalized separately, so $0$ is Nash, $+1$ the optimum, and $-1$ the
worst possible outcome, with the score bounded to $[-1,1]$. A game's score is the
mean over trials (Student-$t$ $95\%$ CI). Tier~1 spans all 13 models across
$k=2$--$10$, with 8--10 games per action count and 12--30 trials each. Not every game
scales to every $k$: battle of the sexes is defined only up to $k=5$, and the cyclic
zero-sum control only for odd $k$, so each action count has six or seven scored games
and two or three controls.

\textbf{Tiers 2--3.} Agents form two teams, and only each team's \emph{composition}
(Section~\ref{sec:methodology}) affects payoffs. Each agent is a separate call
returning a scalar P(A) (Tier~2) or a three-action distribution (Tier~3). Convolving
members' choices (Poisson-binomial for two actions, Poisson-multinomial for three)
gives the team-composition distribution and its expected joint payoff. Teams span two
or three agents (4--6 per game), symmetric and asymmetric, plus a no-dominant-strategy
variant. Team games use an analogous two-sided measure (Eq.~\ref{eq:efficiency}), applied
to expected team welfare in place of individual achieved payoff:
\begin{equation}
s' =
\begin{cases}
\dfrac{a' - N'}{\,O' - N'\,}, & a' \ge N',\\[8pt]
\dfrac{a' - N'}{\,N' - W'\,}, & a' < N',
\end{cases}
\label{eq:team_score}
\end{equation}
with $a'$ the expected joint team welfare, $N'$ the worst symmetric-equilibrium welfare,
$O'$ the best joint-welfare cell, and $W'$ the worst. Constant-sum controls, where every
cell shares the same joint welfare and there is no headroom in either direction, are
excluded. Several archetypes generate zero-sum hybrids whose joint welfare is not
constant and whose worst Nash falls strictly below the optimal cell.

We evaluate 13 models (Table~\ref{tab:model_sources}): two frontier hosted systems, as
an upper bound, and 11 open-weight models from 1B to 14B across five families
(Gemma 2/3/4, Llama 3/3.2, Mistral, DeepSeek-R1, Phi-4), to test whether beating Nash
via self-identity tracks scale, family, or instruction tuning rather than a single
confounded family.

\begin{table}[t]
\caption{Models evaluated.}
\label{tab:model_sources}
\centering
\footnotesize
\begin{tabular}{lll}
\hline
\textbf{Model} & \textbf{Family / Size} & \textbf{Status} \\
\hline
Gemini 3.5 Flash & Google DeepMind & Hosted, frontier \\
GPT-5.6 Luna & OpenAI & Hosted, frontier \\
Mistral 7B & Mistral AI, 7B & Open-weight \\
Gemma 3 12B & Google, 12B & Open-weight \\
Gemma 2 9B & Google, 9B & Open-weight \\
Llama 3.2 3B & Meta, 3B & Open-weight \\
DeepSeek-R1 14B & DeepSeek-AI, 14B & Open-weight, reasoning \\
Phi-4-Mini & Microsoft, 3.8B & Open-weight, compact \\
Llama 3.1 8B & Meta, 8B & Open-weight \\
Gemma 4 e4B & Google, 4B (effective) & Open-weight \\
Gemma 3 4B & Google, 4B & Open-weight \\
Gemma 2 2B & Google, 2B & Open-weight \\
Gemma 3 1B & Google, 1B & Open-weight \\
\hline
\end{tabular}
\end{table}

\section{Results}
\label{sec:results}

\begin{figure}[!t]
    \centering
    \includegraphics[width=0.85\linewidth]{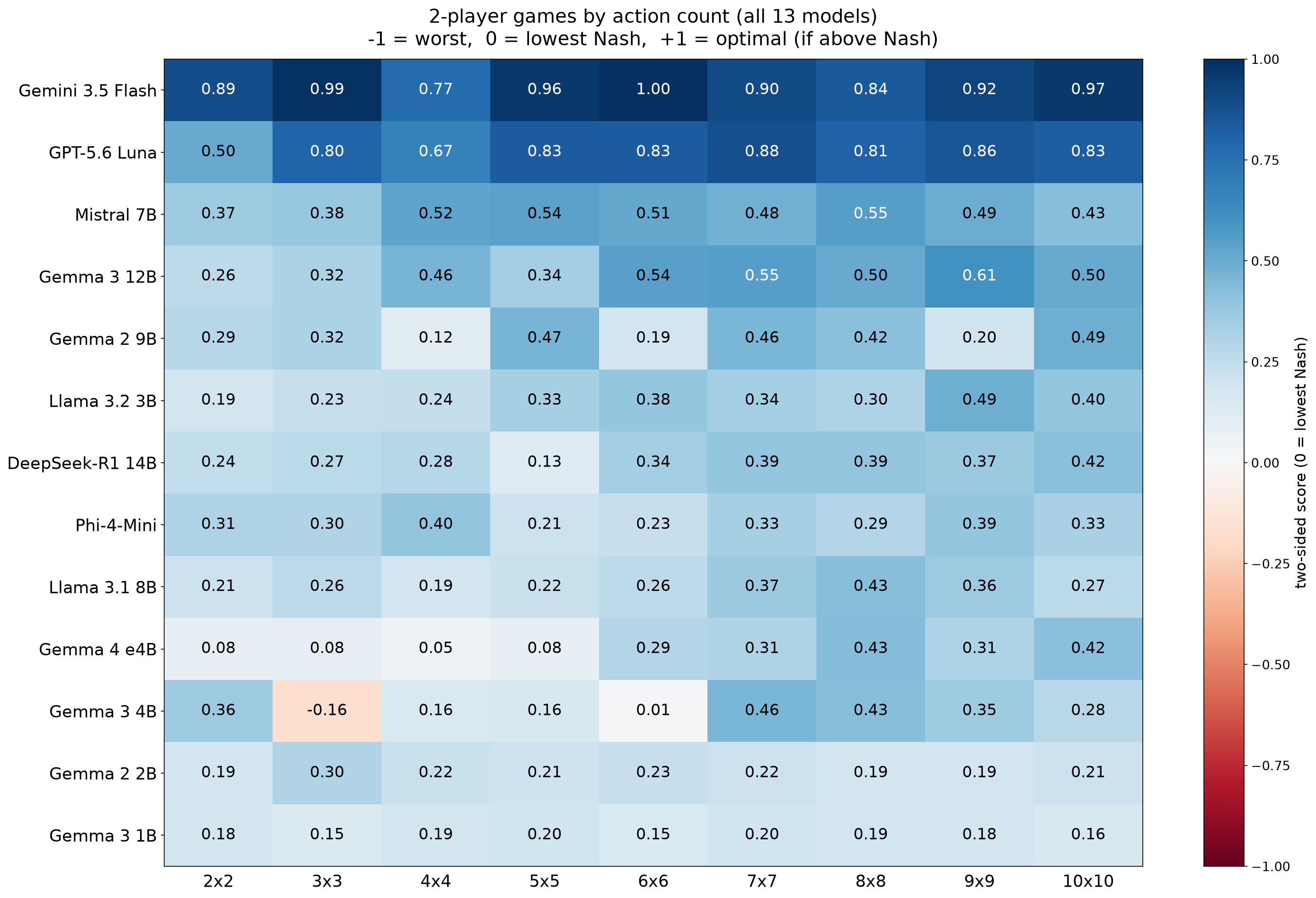}
    \caption{Tier~1 (two-player) games across action count ($k=2$--$10$). $-1=$ worst, $0=$ Nash, $+1=$ optimal.} 
    \label{fig:nash_anchored}
\end{figure}

\begin{figure}[!t]
    \centering
    \includegraphics[width=0.85\linewidth]{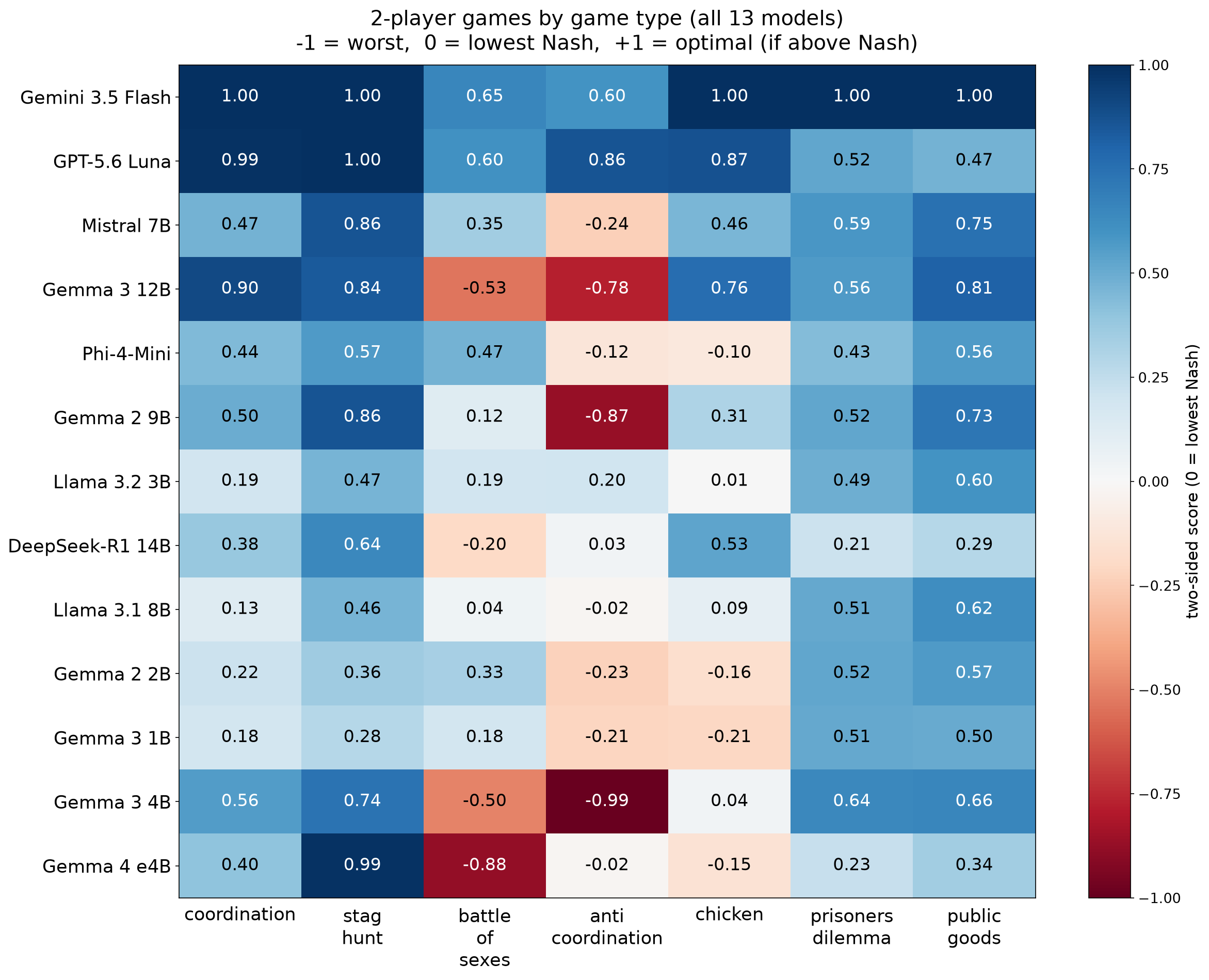}
    \caption{Tier~1 (two-player) across archetypes, averaged over game sizes. $-1=$ worst, $0=$ Nash, $+1=$ optimal.}
    \label{fig:heatmap_archetype}
\end{figure}

\begin{figure}[!t]
    \centering
    \includegraphics[width=0.85\linewidth]{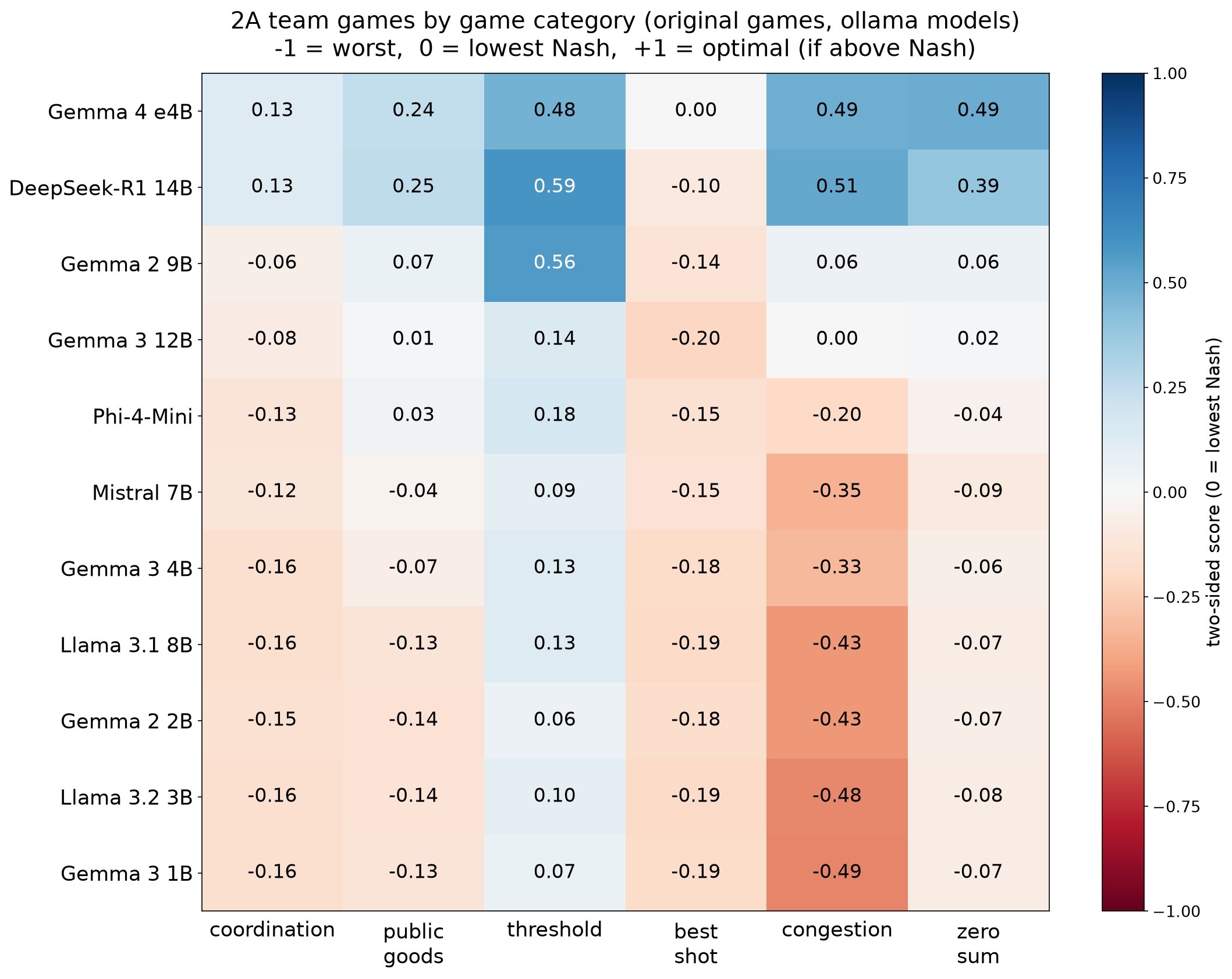}
    \caption{Tier~2 (2-action team) across archetypes. $-1=$ worst, $0=$ Nash, $+1=$ optimal.}
    \label{fig:2_action}
\end{figure}

\begin{figure}[!t]
    \centering
    \includegraphics[width=0.85\linewidth]{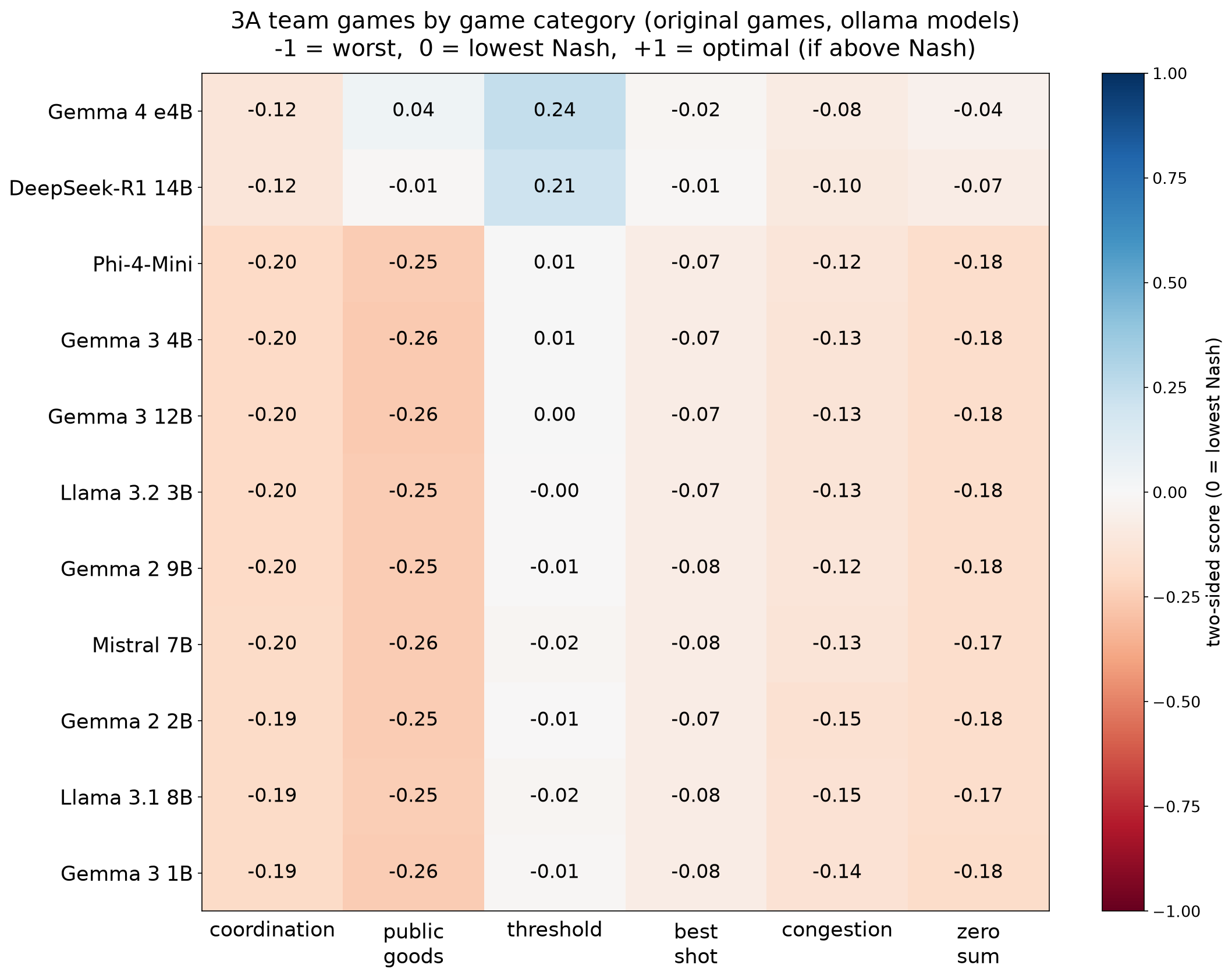}
    \caption{Tier~3 (3-action team) across archetypes. $-1=$ worst, $0=$ Nash, $+1=$ optimal.}
    \label{fig:3_action}
\end{figure}

The scores of Tier 1 are in Figures~\ref{fig:nash_anchored} and~\ref{fig:heatmap_archetype} while Figures~\ref{fig:2_action} and~\ref{fig:3_action}  represent the scores for Tier 2 and 3 respectively.

\setlength{\textfloatsep}{5pt}
\setlength{\intextsep}{5pt}

Across Tier 1, the two frontier hosted models, Gemini 3.5 Flash~\cite{google2026gemini35flash} and GPT-5.6 Luna~\cite{openai2026gpt56}, beat Nash across nearly all $k$ and archetypes. Gemini reaches the optimal joint outcome (1.00) in five of the seven archetypes and GPT-5.6 in four (0.86--1.00), though GPT falls to 0.47--0.60 in prisoner's dilemma, public goods, and battle of the sexes. Open-weight models vary widely~\cite{jiang2023mistral7b,gemmateam2024gemma2,gemmateam2025gemma3,grattafiori2024llama3,meta2024llama32,deepseek2025r1, microsoft2025phi4mini,gemmateam2026gemma4}: Gemma 3 12B and Mistral 7B reach 0.7--0.9 in stag hunt and public goods, while most cluster in 0.2–0.65 in prisoner's dilemma. Chicken scores are far more scattered ($-$0.21 to 0.76), with most models near zero. Performance is largely stable across action count ($k=2$--$10$), indicating that game structure, not action-space size, drives coordination success in the two-player suite.

Archetype (as seen in Figure~\ref{fig:heatmap_archetype}) matters more than model scale. Anti-coordination is hardest: several open-weight models (e.g., Gemma 3 12B, Gemma 2 9B, Gemma 3 4B) score below $-0.5$, near the worst joint cell, despite being positive in coordination and public goods. Battle of the sexes is mixed, with a subset again sharply negative. Chicken is intermediate: most models compress near zero (Gemma 3 12B at 0.76 and DeepSeek-R1 14B at 0.53 are exceptions), while only the hosted models stay strongly positive. Model size does not predict performance: larger local models (e.g., Gemma 3 12B) do not consistently beat smaller ones (e.g., Gemma 3 1B).

The hosted frontier models are omitted from the team tiers for cost reasons: each team game issues one API call per agent per trial (4--10 calls), and running both hosted models over the full team suite was cost-prohibitive.

DeepSeek-R1 14B and Gemma 4 e4B are the clear leaders across nearly every 2-action team archetype (Fig.~\ref{fig:2_action}): DeepSeek-R1 14B tops public goods (0.25), threshold (0.59), and congestion (0.51), while Gemma 4 e4B leads zero-sum (0.49) and is the only model at or above Nash in best shot (0.00). The two tie in coordination (0.13 each). Gemma 2 9B is a distant third overall but rivals the leaders in threshold (0.56). The remaining eight open-weight models score below Nash in coordination, best shot, and congestion (down to $-0.49$), and are mixed in public goods and zero-sum, while threshold is the one archetype scored above Nash for every model (0.06--0.59), making it the most consistently exploitable structure in the 2-action tier.

\setlength{\textfloatsep}{5pt}
\setlength{\intextsep}{5pt}

When the team games move from 2-action to 3-action (Fig.~\ref{fig:3_action}), scores collapse well below Nash across nearly the entire model set. Gemma 4 e4B and DeepSeek-R1 14B retain the only clear positive signal, but it is now concentrated in threshold (0.24 and 0.21, respectively) rather than public goods, which is only marginally positive for Gemma 4 e4B (0.04) and essentially at Nash for DeepSeek-R1 14B ($-0.01$). Their remaining four archetypes are all mildly negative ($-0.01$ to $-0.12$). Every other model scores below Nash in nearly every archetype, clustering between $-0.07$ and $-0.26$, with public goods now the weakest archetype for this group ($-0.25$ to $-0.26$) rather than the sharply negative congestion seen in the 2-action tier. Threshold is the one archetype that stays close to Nash for every model ($-0.02$ to $0.24$). This is a marked collapse from the uniformly strong scores of the 2-action tier (0.06--0.59). Even so, it is the only archetype that does not turn decisively negative.

\section{Discussion}
\label{sec:discussion}

These results suggest that current LLMs can partially exploit knowledge of shared identity, but only in games where doing so requires matching rather than differentiating behavior. In common-interest archetypes (coordination, stag hunt, public goods), identical reasoning is an asset: two copies of the same model tend to converge on the same focal action which pushes joint payoff toward optimal. This is consistent with the intuition of superrationality that motivates our research question in Section~\ref{sec:intro}.

The picture reverses in anti-coordination and, for a subset of models, battle of the sexes, where optimal joint play requires the two agents to select different actions or roles. Because both agents receive symmetric information and reasoning, without any external signal or player index to break the tie, they are more likely to duplicate each other's choice than to spontaneously differentiate. Self-knowledge of shared identity, which helps in common-interest games, becomes a liability here: it offers no mechanism for symmetry-breaking, and can even bias both agents toward the same heuristic. This produces the sharply negative scores observed for several local models.

The hosted--local gap is large in some archetypes (anti-coordination, chicken) but small or reversed in others (coordination, public goods, where Gemma 3 12B and Mistral 7B approach the hosted models), so it is archetype-dependent rather than a uniform capability difference. Because the hosted arms differ from the local arms in both backend and trial count, we cannot cleanly separate model capability from provider-side factors. This absence of a scale trend suggests the capability reflects training data or instruction tuning rather than parameter count.

The team results (Figs.~\ref{fig:2_action}--\ref{fig:3_action}) point to a further limitation: whatever lets a subset of models exploit shared identity in 2-player games does not transfer cleanly to teams, and no model is consistently positive across both tiers. Gemma 4 e4B and DeepSeek-R1 14B lead nearly every 2-action archetype, including congestion (0.49 and 0.51) and zero-sum (0.49 and 0.39). By the 3-action tier both retain a clear positive signal only in threshold (0.24 and 0.21), turning mildly negative in the remaining archetypes. Notably, Gemma 4 e4B was not a standout in the 2-player suite (Fig.~\ref{fig:heatmap_archetype}): it was middling in most archetypes and sharply negative in battle of the sexes ($-0.88$), so its team-coordination ability appears distinct from the capability that drives 2-player self-play. DeepSeek-R1 14B shows the same pattern: strong 2-action performance across nearly all archetypes collapses to a single positive archetype (threshold, 0.21) in the 3-action tier, mirroring Gemma 4 e4B's trajectory and suggesting the loss of signal with added action complexity is a shared limitation rather than a model-specific one.

The action-space sensitivity is stark: Gemma 4 e4B's scores across archetypes fall from a range of $0.00$ to $0.49$ in the 2-action suite (Fig.~\ref{fig:2_action}) to $-0.12$ to $0.24$ in the 3-action suite (Fig.~\ref{fig:3_action}). The coordination archetype specifically reverses sign, from $0.13$ (above Nash) to $-0.12$ (below Nash) -- a reversal of its coordination advantage. This contrasts with the stable 2-player performance across action counts noted above (Fig.~\ref{fig:nash_anchored}). A plausible explanation is combinatorial: with a single partner an agent need only match one other reasoner regardless of the action count, whereas on a team each added action multiplies the possible team compositions it must reason over (Section~\ref{sec:methodology}), making beliefs about how teammates split their choices harder to form.

Each two-player game was played 12 to 30 times and each team game 8 to 9 times, always with independent seeds and rotated action labels, amounting to around 53{,}000 game instances.

\section{Conclusion}
We introduced a benchmark of one-shot, no-communication games to test whether independent copies of the same LLM can coordinate purely by reasoning about a counterpart known to share their identity. In two-player games, both hosted models and several open-weight models scored above Nash in most archetypes, approaching the optimal joint outcome, though this reversed in anti-coordination and battle of the sexes in open-weight models. Performance showed no clear decline as the number of actions increased, and model scale did not predict success. In team games coordination was far less reliable: most models scored below Nash in the 3-action tier, and even the strongest 2-action performers, Gemma 4 e4B and DeepSeek-R1 14B, retained a positive signal only in threshold once the action space grew. Overall, copies of the same LLM can coordinate above the Nash baseline without communication, especially when success requires matching actions, but this ability is archetype-dependent and degrades sharply in larger, higher-dimensional team settings.

\section*{Acknowledgment}
Fig.~\ref{fig:bos_example} was created using OpenAI image-generation tools and reviewed by the authors.


\begin{thebibliography}{00}
\bibitem{cao1997cooperative}
Y.~U. Cao, A.~S. Fukunaga, and A.~B. Kahng, ``Cooperative mobile robotics:
Antecedents and directions,'' \textit{Autonomous Robots}, vol.~4, no.~1,
pp.~7--27, 1997, doi: 10.1023/A:1008855018923.
\bibitem{gerkey2004taskallocation}
B.~P. Gerkey and M.~J. Matari\'{c}, ``A formal analysis and taxonomy of task
allocation in multi-robot systems,'' \textit{The International Journal of
Robotics Research}, vol.~23, no.~9, pp.~939--954, 2004,
doi: 10.1177/0278364904045564.
\bibitem{ismayilov2025decentralized} M. Ismayilov, E. Meriaux, S. Wen, and G. Dudek, ``Decentralized multi-agent goal assignment for path planning using large language models,'' in \emph{2025 IEEE MIT Undergraduate Research Technology Conference (URTC)}, 2025, pp. 1--5.
\bibitem{akata2025repeated}
E.~Akata, L.~Schulz, J.~Coda-Forno, S.~J. Oh, M.~Bethge, and E.~Schulz,
``Playing repeated games with large language models,'' \textit{Nature Human
Behaviour}, vol.~9, pp.~1380--1390, 2025.
\bibitem{agashe2025coordination}
S.~Agashe, Y.~Fan, A.~Reyna, and X.~E. Wang, ``LLM-Coordination: Evaluating
and analyzing multi-agent coordination abilities in large language models,''
in \textit{Findings of the Association for Computational Linguistics: NAACL
2025}, pp.~8053--8072, 2025, doi: 10.18653/v1/2025.findings-naacl.448.
\bibitem{nash1950equilibrium}
J.~F. Nash, ``Equilibrium points in $n$-person games,''
\textit{Proceedings of the National Academy of Sciences}, vol.~36, no.~1,
pp.~48--49, 1950.
\bibitem{aumann1974correlated}
R.~J. Aumann, ``Subjectivity and correlation in randomized strategies,''
\textit{Journal of Mathematical Economics}, vol.~1, no.~1, pp.~67--96, 1974.
\bibitem{schelling1960strategy}
T.~C. Schelling, \textit{The Strategy of Conflict}.
Cambridge, MA: Harvard University Press, 1960.
\bibitem{lewis1969convention}
D.~K. Lewis, \textit{Convention: A Philosophical Study}.
Cambridge, MA: Harvard University Press, 1969.
\bibitem{hofstadter1983superrationality}
D.~R. Hofstadter, ``Dilemmas for superrational thinkers, leading up to a
luring lottery,'' \textit{Scientific American}, vol.~248, no.~6, Jun.~1983.
\bibitem{tennenholtz2004program}
M.~Tennenholtz, ``Program equilibrium,'' \textit{Games and Economic
Behavior}, vol.~49, no.~2, pp.~363--373, 2004.
\bibitem{hu2020otherplay}
H.~Hu, A.~Lerer, A.~Peysakhovich, and J.~Foerster, ``Other-play for
zero-shot coordination,'' in \textit{Proc. Int. Conf. Machine Learning
(ICML)}, 2020, pp.~4399--4408.
\bibitem{duan2024gtbench}
J.~Duan, R.~Zhang, J.~Diffenderfer, B.~Kailkhura, L.~Sun, E.~Stengel-Eskin,
M.~Bansal, T.~Chen, and K.~Xu, ``GTBench: Uncovering the strategic reasoning
capabilities of LLMs via game-theoretic evaluations,'' in \textit{Advances in
Neural Information Processing Systems (NeurIPS)}, 2024.
\bibitem{lemkehowson1964}
C.~E. Lemke and J.~T. Howson, Jr., ``Equilibrium points of bimatrix
games,'' \textit{Journal of the Society for Industrial and Applied
Mathematics}, vol.~12, no.~2, pp.~413--423, 1964,
doi: 10.1137/0112033.
\bibitem{google2026gemini35flash}
Google DeepMind, ``Gemini 3.5 Flash: Model card,'' May 2026. [Online].
Available:
\url{https://deepmind.google/models/model-cards/gemini-3-5-flash/}.
\bibitem{openai2026gpt56}
OpenAI, ``GPT-5.6 system card,'' Jul. 2026. [Online]. Available:
\url{https://deploymentsafety.openai.com/gpt-5-6}.
\bibitem{jiang2023mistral7b}
A.~Q. Jiang \textit{et al.}, ``Mistral 7B,'' arXiv:2310.06825, 2023,
doi: 10.48550/arXiv.2310.06825.
\bibitem{gemmateam2024gemma2}
Gemma Team, ``Gemma 2: Improving open language models at a practical
size,'' arXiv:2408.00118, 2024,
doi: 10.48550/arXiv.2408.00118.
\bibitem{gemmateam2025gemma3}
Gemma Team, ``Gemma 3 technical report,'' arXiv:2503.19786, 2025,
doi: 10.48550/arXiv.2503.19786.
\bibitem{grattafiori2024llama3}
A.~Grattafiori \textit{et al.}, ``The Llama 3 herd of models,''
arXiv:2407.21783, 2024,
doi: 10.48550/arXiv.2407.21783.
\bibitem{meta2024llama32}
Meta AI, ``Llama 3.2: Revolutionizing edge AI and vision with open,
customizable models,'' Sep. 2024. [Online]. Available:
\url{https://ai.meta.com/blog/llama-3-2-connect-2024-vision-edge-mobile-devices/}.
\bibitem{deepseek2025r1}
DeepSeek-AI, ``DeepSeek-R1: Incentivizing reasoning capability in LLMs
via reinforcement learning,'' arXiv:2501.12948, 2025,
doi: 10.48550/arXiv.2501.12948.
\bibitem{microsoft2025phi4mini}
Microsoft, ``Phi-4-Mini technical report: Compact yet powerful
multimodal language models via mixture-of-LoRAs,''
arXiv:2503.01743, 2025,
doi: 10.48550/arXiv.2503.01743.
\bibitem{gemmateam2026gemma4}
Gemma Team, ``Gemma 4 technical report,'' arXiv:2607.02770, 2026,
doi: 10.48550/arXiv.2607.02770.
\end{thebibliography}
\end{document}